\documentclass[pdflatex,sn-mathphys-num]{sn-jnl}% Math and Physical Sciences Numbered Reference Style
\usepackage{graphicx}%
\usepackage{multirow}%
\usepackage{amsmath,amssymb,amsfonts}%
\usepackage{amsthm}%
\usepackage{mathrsfs}%
\usepackage[title]{appendix}%
\usepackage{xcolor}%
\usepackage{textcomp}%
\usepackage[utf8]{inputenc}%
\usepackage{manyfoot}%
\usepackage{booktabs}%
\usepackage{algorithm}%
\usepackage{algorithmicx}%
\usepackage{algpseudocode}%
\usepackage{listings}%
\usepackage{fancyhdr}
\usepackage{subcaption}

\theoremstyle{thmstyleone}%
\theoremstyle{thmstyletwo}%

\theoremstyle{thmstylethree}%

\begin{document}

\makeatletter
\newcommand{\authorrowbreak}{%
  \g@addto@macro\artauthors{%
    \global\advance\punctcount by -1%
    \par
    \let\authorsep\@empty
  }%
}
\makeatother

\title[CCI as Human--Environment Belief Updating]{Uncovering Salience-Driven Dynamics in Consumer Confidence with Generative Social Simulation}

%%=============================================================%%
%% GivenName	-> \fnm{Joergen W.}
%% Particle	-> \spfx{van der} -> surname prefix
%% FamilyName	-> \sur{Ploeg}
%% Suffix	-> \sfx{IV}
%% \author*[1,2]{\fnm{Joergen W.} \spfx{van der} \sur{Ploeg} 
%%  \sfx{IV}}\email{iauthor@gmail.com}
%%=============================================================%%

\author[1,2]{\fnm{Yixu} \sur{Huang}}
\equalcont{These authors contributed equally to this work.}

\author[1]{\fnm{Yunlu} \sur{Yin}}
\equalcont{These authors contributed equally to this work.}

\author[1,2]{\fnm{Jiayu} \sur{Lin}}
\authorrowbreak

\author[1,2]{\fnm{Xinnong} \sur{Zhang}}

\author[1,2]{\fnm{Jia} \sur{Wang}}

\author[3]{\fnm{Siyuan} \sur{Wang}}
\authorrowbreak

\author[1,2]{\fnm{Xuanjing} \sur{Huang}}

\author*[1]{\fnm{Liyin} \sur{Jin}}
\email{jinliyin@fudan.edu.cn}

\author*[1,2]{\fnm{Zhongyu} \sur{Wei}}
\email{zywei@fudan.edu.cn}

\affil*[1]{\orgname{Fudan University}}

\affil[2]{\orgname{Shanghai Innovation Institute}}

\affil[3]{\orgname{The Chinese University of Hong Kong}}

%%==================================%%
%% Sample for unstructured abstract %%
%%==================================%%

\abstract{Consumer confidence is typically modeled as a persistent macroeconomic index, yet its movements arise from households that interpret economic information through heterogeneous constraints, exposures, prior beliefs, and attention. We introduce ConsumerSim, a generative Human--Environment response framework that reconstructs Consumer Confidence Index (CCI) dynamics from a microdata-calibrated synthetic population, time-stamped macroeconomic, financial, policy, and news signals, survey-like response generation, post-stratified belief expansion, and behavioral inertia alignment. Across U.S., EU27, and Japanese official CCI target series, ConsumerSim ranks first among persistence, time-series, regression, and information-augmented baselines on the reported reconstruction metrics, with clear gains around high-salience shocks. Its reconstructed signal also improves short-horizon prediction of real activity, most consistently for housing outcomes. Mechanism analyses show that CCI movements concentrate around salient events; subgroup trajectories often align in direction while differing in magnitude; and signal sensitivity varies across income, homeownership, education, and political-alignment groups. Population-expansion and ablation results indicate that representative aggregation, situational signals, persona heterogeneity, and inertia are necessary for both accuracy and diagnosis. The findings support a behavioral view of consumer confidence as an interpretable Human--Environment response process rather than a purely aggregate time series.}


%%\pacs[JEL Classification]{D8, H51}

%%\pacs[MSC Classification]{35A01, 65L10, 65L12, 65L20, 65L70}

\maketitle

\section{Introduction}\label{sec1}

Consumer confidence surveys measure how households assess their own financial prospects and the broader economy~\citep{katona1951psychological,katona1975psychological,dominitz2004measure,manski2004measuring}. Due to the role that households' current financial conditions and expectations about future conditions play in consumption decisions, these confidence judgments matter for understanding consumer behavior~\citep{souleles2001consumer,howrey2001predictive,carroll2003macroeconomic,ludvigson2004consumer,barsky2012information}. For this reason, many economies track these judgments through official or widely used Consumer Confidence Index (CCI) survey systems, including the U.S. CCI target based on the University of Michigan Surveys of Consumers~\citep{michigan2026surveys,surveys2024technical}, the EU27 CCI series based on European Commission/Eurostat harmonized survey balances~\citep{europeancommission2025bcs,eurostat2026businessconsumer,goldrian2001evaluation}, and the Japan CCI series based on the Cabinet Office/ESRI survey series~\citep{cabinetoffice2026consumerconfidence}. These CCI~\footnote{For consistency, we use CCI as the shorthand for these regional official target series throughout the paper.} series aggregate these subjective judgments and have long been linked to later spending and saving behavior~\citep{katona1951psychological,katona1975psychological,howrey2001predictive,ludvigson2004consumer}.

Yet CCI does not behave like a smoothly evolving macroeconomic variable. It is built from categorical judgments made by households with different circumstances, such as household income and exposure to housing costs, under changing information environments, such as inflation news or policy changes. The aggregate series is often persistent in quiet periods but can shift sharply when public information becomes broadly relevant. Although prior work links CCI to macroeconomic outcomes and demographic expectation differences~\citep{carroll2003macroeconomic,ludvigson2004consumer,souleles2004expectations,manski2004measuring}, forecasting models usually treat it as an aggregate series predicted from lags, macro-financial variables, survey expectations, or text sentiment~\citep{box2008time,stock2007why,shapiro2022measuring,kim2023forecasting}. This aggregate framing obscures the diagnostic question: \emph{how do household circumstances and environmental information combine to produce collective CCI movement?} Aggregate CCI records how far the index moved, but it suppresses which households updated, which signals reached them, and whether those updates were broad enough to persist.

Social simulation provides a methodological starting point for this reconstruction, explaining aggregate patterns as outcomes of heterogeneous agents interacting with structured environments~\citep{epstein1996growing,grimm2006standard,gao2024large,zhang2025socioverse}. A central challenge is therefore to reconstruct the latent Human--Environment process beneath the index. Inspired by field theory, which treats behavior as shaped by the person and the environment~\citep{lewin1936principles}, we represent this process not as a deterministic scalar function but as a response distribution. Conceptually, for persona $i$ in time $t$ and survey question $q$, $\mathbf{p}_{i,t,q}=\mathcal{R}(h_i,s_t,q)$, where $h_i$ is the household persona, $s_t$ is the situational signal field, and $\mathcal{R}$ is the response-generation process that maps the person and environment into probabilities over survey response categories. This process includes the dynamic matching and salience assignment through which a shared economic environment becomes personally relevant to different households. CCI can therefore be viewed not simply as a number to be forecast from past values but as an aggregate of probabilistic household responses formed under changing macroeconomic, financial, policy, and news conditions. Since the reconstruction should stay anchored to observed CCI dynamics, forecasting accuracy remains important. The broader objective is to use that reconstruction to decompose a CCI movement into its population composition, signal composition, and timing.

\begin{figure}[t]
    \centering
    \includegraphics[
        page=1,
        trim=0cm 0cm 0cm 0cm,
        clip=true,
        width=0.9\linewidth
        ]{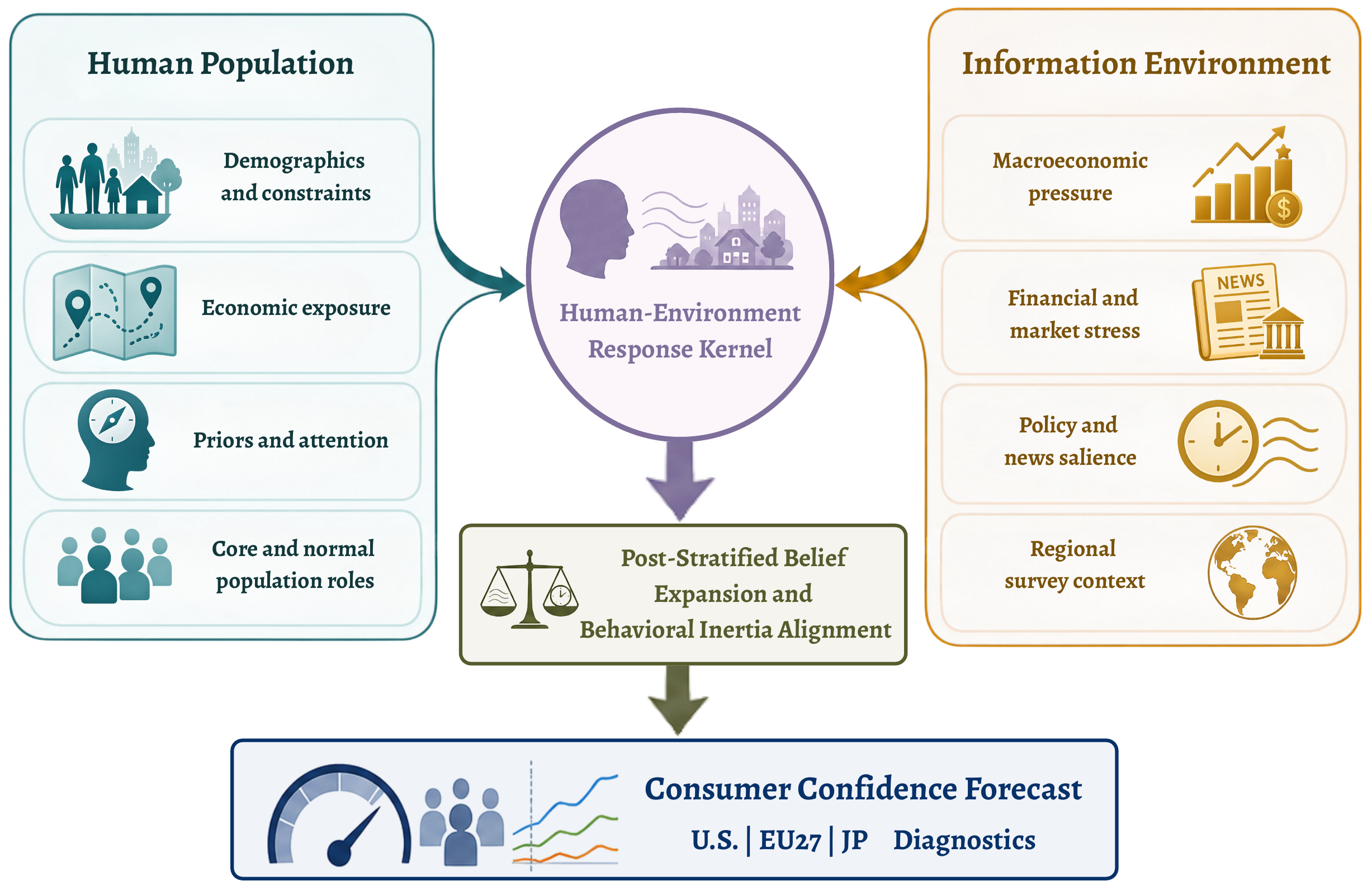}
    \vspace{0.5em}
    \caption{Human--Environment response framework of ConsumerSim. The Human Population module builds heterogeneous consumers through life position, exposure, behavioral orientation, and core/normal population roles; the Information Environment module builds a situational signal field with economic pressure, public salience, and dynamic matching. Their interaction is mapped through a Human--Environment Response Kernel, expanded through population aggregation, aligned with behavioral inertia, and converted into CCI forecasts and diagnostic outputs.}
    \label{fig:hnb_framework}
\end{figure}

Therefore, we introduce ConsumerSim, a generative-agent simulation framework that reconstructs CCI as a Human--Environment response process (Figure~\ref{fig:hnb_framework}). The framework has five key design choices. First, it constructs a microdata-calibrated population of synthetic households so that CCI movements can be traced to life position, exposure, and behavioral orientation. Second, it builds monthly information environments that organize different types of signals by salience and relevance. Third, it uses a human--environment response kernel to generate survey-like CCI responses. Fourth, it expands responses from a fixed high-resolution core population to a broader normal population through post-stratified belief expansion, so the aggregate remains representative rather than merely reflecting the core agents. Fifth, it aligns the expanded aggregate with observed behavioral inertia, recognizing that public expectations are sticky and should not be reset each month. The resulting response shares are then converted into official-style CCI measures and diagnostic outputs. These choices turn the idea of CCI movement as a salience-triggered, heterogeneous belief-updating process into a concrete model. Households differ in socioeconomic position and economic exposure, while environments differ in which signals become salient to whom. Because people do not absorb all information equally, attention, heuristics, and lived experience shape what they notice and how strongly they update~\citep{tversky1974judgment,lord1979biased,nickerson1998confirmation,rollwage2020confidence,bordalo2020overreaction,bordalo2022salience,das2020socioeconomic,malmendier2016learning,coibion2015information}.

We evaluate ConsumerSim through two validation checks and a set of diagnostic analyses. The validation checks establish that the simulated response process is sufficiently close to official CCI ground truth to support interpretation: ConsumerSim ranks first across the reported United States, European, and Japanese evaluations, with representative MAE/RMSE values of 3.45/4.56, 1.607/2.737, and 1.363/2.040, respectively~\citep{michigan2026surveys,surveys2024technical,europeancommission2025bcs,eurostat2026businessconsumer,goldrian2001evaluation,cabinetoffice2026consumerconfidence}. A downstream validation further shows that the reconstructed signal is not merely numerically close to official CCI. Replacing official CCI with ConsumerSim-predicted CCI improves near-term housing prediction, with positive incremental $R^2$ and lower RMSE across short forecast horizons.

Having established that the reconstructed series is empirically anchored, we then use ConsumerSim for its central purpose: diagnosing how confidence forms and changes. The diagnostic analyses show that CCI shifts concentrate around salient shocks rather than being evenly distributed over time. These shocks can align subgroup trajectories while preserving heterogeneous magnitudes by income, homeownership, education, and political alignment. The sensitive groups also vary by signal type: inflation, housing, financial stress, labor risk, and trade policy each reach different parts of the population with different strength. Within-month diagnostic paths further localize when newly visible information enters the confidence process. Finally, mechanism ablations show that removing post-stratified belief expansion, Behavioral Inertia Alignment, the Situational Signal Field, or the Hierarchical Persona Substrate weakens both aggregate reconstruction and mechanism-level diagnosis.

Together, these findings support the Human--Environment interpretation: consumer confidence moves collectively when salient environments align belief updating, but the size and timing of updates remain structured by household exposure, priors, and attention. The contribution is therefore not a replacement survey or a stand-alone forecasting tool, but a modular diagnostic system that separates population construction, information-environment representation, response generation, and aggregate calibration. This system can answer which households are moving, which signals are reaching them, whether the movement is broad or narrow, and whether it is likely to persist. It thereby turns the same headline CCI movement into a more actionable diagnosis while remaining extensible to other expectation-driven domains.

% 诊断的结论在 introduction 中点出
% 补充可控实验
% 补一个 probing 证明 llm 没有记下来知识
% 补充 lambda ()

\section{Results}\label{sec2}

\subsection{ConsumerSim reconstructs CCI series and predicts subsequent real activity.}

\begin{figure}[t]
\centering
\includegraphics[
page=1,
trim=0cm 0cm 0cm 0cm,
clip=true,
width=\linewidth
]{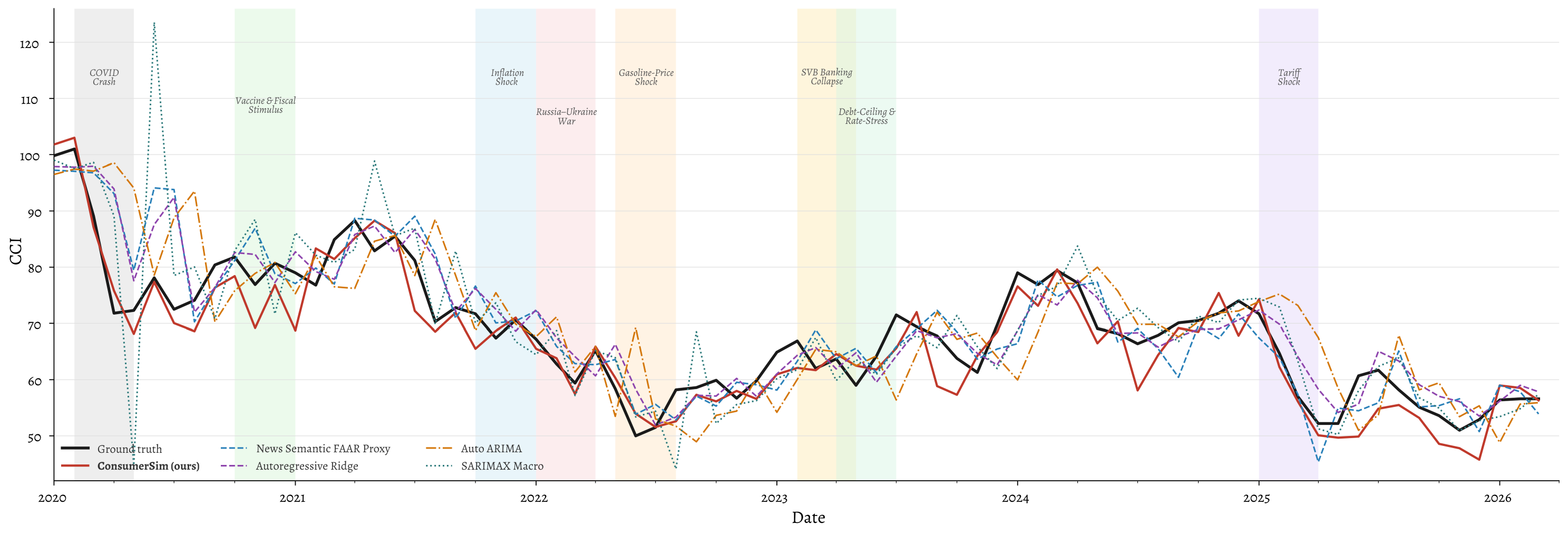}
\caption{CCI shifts and model performance in high-salience regimes. The official U.S. CCI series is shown with selected salient shocks shaded. Observed values are plotted alongside ConsumerSim and representative baseline models. Baseline methods lag during abrupt CCI shifts, whereas ConsumerSim more closely tracks the timing and magnitude of realized CCI movements.}
\label{fig:cci_temporal_comparison}
\end{figure}

\paragraph{Cross-regional reconstruction.} ConsumerSim accurately reconstructs CCI series across regions (Table~\ref{tab:cci_forecast_comparison}). We evaluate three regional target series on equal footing: the U.S. CCI target based on the University of Michigan Surveys of Consumers, the EU27 CCI series based on Eurostat harmonized survey balances, and the Japan CCI series based on the Cabinet Office/ESRI survey series. Target definitions and evaluation settings are summarized in Appendix~\ref{app:cross_regional_targets}. Across all three targets, ConsumerSim ranks first on the reported metrics. It achieves MAE/RMSE of 3.45/4.56 for the U.S., 1.607/2.737 for EU27, and 1.363/2.040 for Japan, with Pearson/Spearman correlations of 0.9590/0.9519, 0.882/0.849, and 0.824/0.840, respectively. The U.S. time-series comparison in Figure~\ref{fig:cci_temporal_comparison} shows that these reconstruction gains are especially visible around sharp CCI movements. Because the U.S. battery includes forward-looking business-expectation items (BUS12 and BUS5), Appendix~\ref{app:post_cutoff_us_robustness} reports a U.S.-only post-cutoff check for April--June 2026, outside the March 2026 GPT-4o knowledge window. These results establish that the reconstructed series closely track official CCI dynamics across independent measurement systems before the mechanism analyses below examine where the gains are concentrated.

\begin{table*}[t]
\centering
\caption{CCI forecasting performance for the U.S., EU27, and Japan over January 2020--March 2026, spanning 75 monthly observations. Lower MAE and RMSE indicate better level accuracy; higher Pearson $r$ and Spearman $\rho$ indicate stronger co-movement with the official CCI series. Best available results within each region and metric are shown in bold; arrows indicate the preferred direction ($\downarrow$ lower is better, $\uparrow$ higher is better).}
\label{tab:cci_forecast_comparison}
\scriptsize
\resizebox{\textwidth}{!}{%
\begin{tabular}{@{}lcccccccccccc@{}}
\toprule
\multirow{2}{*}{\textbf{Target / method}} &
\multicolumn{3}{c}{\textbf{MAE} $\downarrow$} &
\multicolumn{3}{c}{\textbf{RMSE} $\downarrow$} &
\multicolumn{3}{c}{\textbf{Pearson $r$} $\uparrow$} &
\multicolumn{3}{c}{\textbf{Spearman $\rho$} $\uparrow$}\\
\cmidrule(lr){2-4}\cmidrule(lr){5-7}\cmidrule(lr){8-10}\cmidrule(l){11-13}
& \textbf{U.S.} & \textbf{EU27} & \textbf{Japan}
& \textbf{U.S.} & \textbf{EU27} & \textbf{Japan}
& \textbf{U.S.} & \textbf{EU27} & \textbf{Japan}
& \textbf{U.S.} & \textbf{EU27} & \textbf{Japan}\\
\midrule
\textbf{ConsumerSim} & \textbf{3.45} & \textbf{1.607} & \textbf{1.363 }& \textbf{4.56} & \textbf{2.737} & \textbf{2.040} & \textbf{0.9590} & \textbf{0.882} & \textbf{0.824} & \textbf{0.9519} & \textbf{0.849} & \textbf{0.840}\\
% Naive & 3.55 & 1.609 & \textbf{1.352} & 4.61 & \textbf{2.721} & 2.051 & 0.9567 & 0.881 & 0.818 & 0.9417 & 0.845 & 0.831\\
AR Ridge & 3.77 & 2.152 & 1.393 & 5.18 & 4.205 & 2.295 & 0.9459 & 0.749 & 0.794 & 0.9363 & 0.806 & 0.839\\
Expectations Reg. & 3.96 & 2.490 & 1.643 & 5.45 & 4.082 & 2.774 & 0.9443 & 0.834 & 0.760 & 0.9346 & 0.799 & 0.804\\
News Semantic FAAR & 4.02 & 2.380 & 1.550 & 5.63 & 3.769 & 2.374 & 0.9378 & 0.791 & 0.753 & 0.9331 & 0.797 & 0.816\\
3-Month Mean & 4.34 & 2.216 & 1.816 & 5.88 & 3.422 & 2.722 & 0.9295 & 0.804 & 0.652 & 0.9092 & 0.782 & 0.729\\
SARIMAX Macro & 4.93 & 3.072 & 2.854 & 7.98 & 5.191 & 4.932 & 0.8933 & 0.755 & 0.609 & 0.9082 & 0.735 & 0.648\\
ETS Damped Trend & 4.96 & 1.833 & 1.492 & 6.94 & 3.471 & 2.438 & 0.9058 & 0.830 & 0.798 & 0.8767 & 0.833 & 0.833\\
Theta Model & 5.08 & 1.681 & 1.374 & 6.95 & 2.848 & 2.063 & 0.9030 & 0.870 & 0.818 & 0.8775 & 0.840 & 0.830\\
Auto ARIMA & 5.44 & 2.373 & 1.642 & 7.76 & 4.173 & 2.576 & 0.8852 & 0.767 & 0.778 & 0.8707 & 0.775 & 0.775\\
Prophet Macro & 5.89 & 4.148 & 2.629 & 8.10 & 5.664 & 3.664 & 0.8985 & 0.564 & 0.541 & 0.8965 & 0.384 & 0.596\\
Prophet Univ. & 6.06 & 4.806 & 2.801 & 7.96 & 6.175 & 3.596 & 0.8714 & 0.274 & 0.468 & 0.8277 & 0.321 & 0.555\\
Macro-Market-News Ridge & 6.21 & 2.511 & 1.951 & 14.40 & 3.810 & 2.841 & 0.7215 & 0.836 & 0.783 & 0.8758 & 0.780 & 0.763\\
Macro Stable Ridge & 6.45 & 2.490 & 1.959 & 15.12 & 4.082 & 2.997 & 0.6893 & 0.834 & 0.771 & 0.8705 & 0.799 & 0.783\\
\bottomrule
\end{tabular}
}
\footnotetext{U.S. = United States CCI series; EU27 = European Union 27-country CCI series based on harmonized Eurostat survey balances; Japan = CCI series based on the Cabinet Office/ESRI survey series. All reported regional evaluations cover January 2020--March 2026. Method abbreviations: AR = autoregressive; Reg. = regression; FAAR = factor-augmented autoregressive; SARIMAX = seasonal autoregressive integrated moving average with exogenous regressors; ETS = exponential smoothing; Auto ARIMA = automatic autoregressive integrated moving average. Best values are bolded within each region and metric.}
\end{table*}

\begin{figure}[t]
\centering
\includegraphics[width=\linewidth]{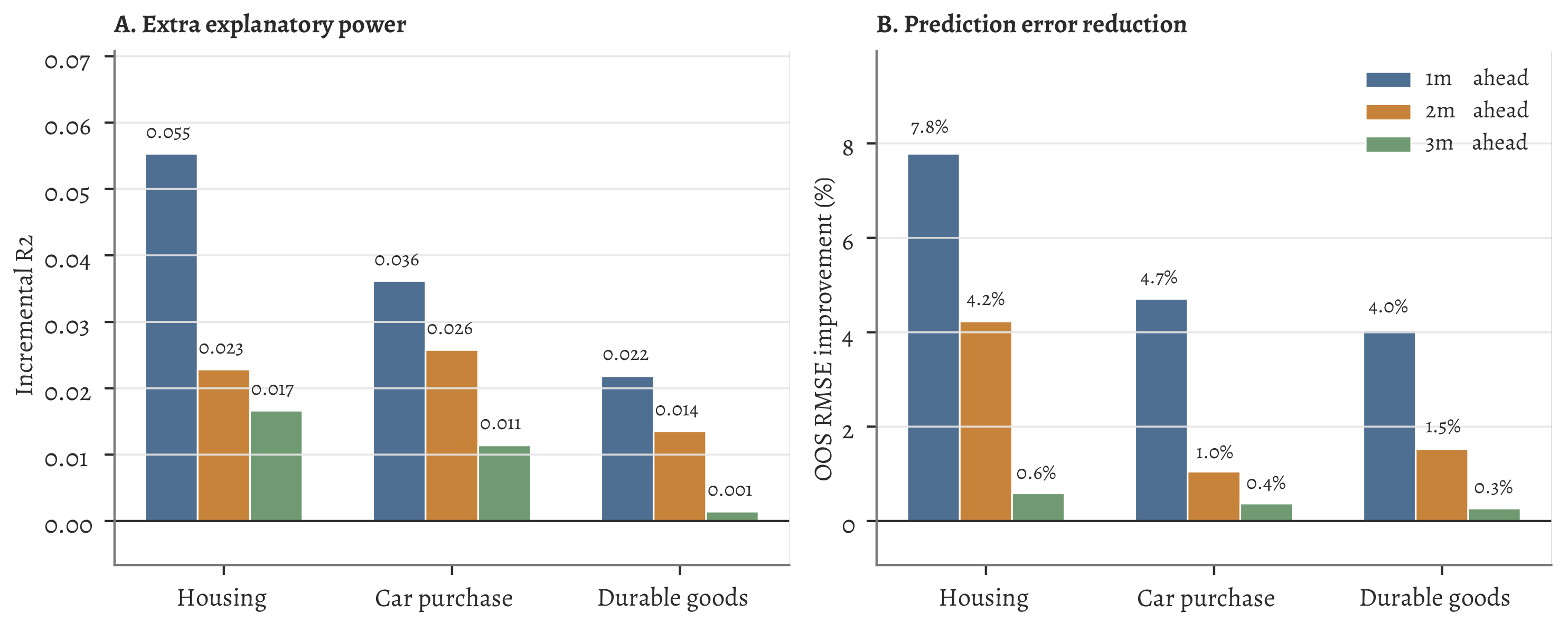}
\caption{Downstream real-activity validation. Models are fit on April 2025--March 2026; the one-, two-, and three-month-ahead forecasts target April, May, and June 2026. Panel A reports incremental $R^2$ from replacing official CCI with ConsumerSim-predicted CCI; Panel B reports RMSE improvement over the official-CCI specification.}
\label{fig:activity_growth_predictions}
\end{figure}

\paragraph{Real-activity prediction.} ConsumerSim is also useful beyond reconstructing CCI because its reconstructed signal improves short-horizon prediction of real economic activity. The validation compares otherwise matched predictive models that differ only in whether they use official CCI or ConsumerSim-predicted CCI as the confidence input; the full regression setup, horizons, ConsumerSim--official-CCI gap definition, and evaluation metrics are reported in Appendix~\ref{app:real_activity_validation}. The activity outcomes cover three consumption-related domains: new home sales, housing starts, and building permits in the housing block; light-vehicle sales in the vehicle block; and real durable-goods consumption expenditure in the durable-goods block. Across these candidate outcomes, housing provides the clearest evidence of incremental predictive value (Figure~\ref{fig:activity_growth_predictions}). Replacing official CCI with ConsumerSim-predicted CCI yields positive housing gains at all horizons, with incremental $R^2$ of 0.055, 0.023, and 0.017 at one-, two-, and three-month horizons and corresponding out-of-sample RMSE reductions of 7.8\%, 4.2\%, and 0.6\%. The gains are also positive for car purchases and durable goods, with the strongest car-purchase gains at the one- and two-month horizons and the strongest durable-goods gains at the one-month horizon, indicating that the ConsumerSim--official-CCI gap is most informative for expectation-sensitive housing activity while also carrying weaker downstream signal for other consumption-related domains.

\subsection{CCI shifts concentrate around salient shocks.}

We next examine whether CCI shifts cluster around salient shocks. We define salient shocks as periods when macroeconomic, financial, geopolitical, or policy information becomes unusually visible and directly relevant to household economic expectations. In the U.S. series, selected salient movements occur around the COVID-19 crash, the inflation shock, the Russia--Ukraine war outbreak, the Silicon Valley Bank (SVB) banking collapse, and the tariff shock (Figure~\ref{fig:cci_temporal_comparison}). Baseline methods that rely more heavily on persistence or smooth temporal dynamics track gradual movements but lag around abrupt changes. ConsumerSim remains closer to the realized series in these windows, consistent with the idea that an event-aware information environment is needed to reconstruct sharp movements in CCI.

We quantify this pattern by evaluating model performance around annotated high-salience regimes (Table~\ref{tab:shock_robustness}). Each row reports a central event month, and performance is measured over the three-month window spanning the preceding month, the event month, and the following month. ConsumerSim ranks first in the combined high-salience-window evaluation and ranks first across all three reported metrics in most individual shock windows. Across the full series, persistence remains a strong baseline; the relative advantage of ConsumerSim becomes most informative when CCI departs from smooth extrapolation. This supports the first mechanism implied by the framework: CCI shifts are concentrated in regimes where updating is event-driven rather than continuously distributed over time.

\begin{table*}[t]
\centering
\caption{Robustness of forecasting performance around high-salience economic regimes. Rankings indicate relative performance among 14 competing methods (\#1 = best). ConsumerSim ranks first in the combined high-salience-window evaluation and in most individual event windows; non-first ranks are shown explicitly.}
\label{tab:shock_robustness}
\scriptsize
\resizebox{0.8\textwidth}{!}{%
\begin{tabular}{@{}p{0.48\textwidth}c c c@{}}
\toprule
\textbf{Regime} & \shortstack{\textbf{Event}\\\textbf{window}} & \shortstack{\textbf{MAE}\\\textbf{rank}} & \shortstack{\textbf{RMSE}\\\textbf{rank}} \\
\midrule
Oil-Price Collapse & 2014-12 & \#1/14 & \#1/14 \\
Greek Debt-Crisis Escalation & 2015-07 & \#1/14 & \#2/14 \\
China Devaluation and Global Selloff & 2015-08 & \#1/14 & \#1/14 \\
Federal Reserve Liftoff & 2015-12 & \#1/14 & \#1/14 \\
Global Equity and Oil Selloff & 2016-01 & \#1/14 & \#1/14 \\
Brexit Referendum & 2016-06 & \#1/14 & \#1/14 \\
U.S. Presidential Election & 2016-11 & \#2/14 & \#1/14 \\
Hurricane Harvey--Irma Disruption & 2017-09 & \#1/14 & \#1/14 \\
Market Volatility Spike & 2018-02 & \#1/14 & \#1/14 \\
Government Shutdown and Market Selloff & 2018-12 & \#2/14 & \#3/14 \\
Trade-War Escalation & 2019-08 & \#1/14 & \#1/14 \\
COVID Crash & 2020-03 & \#1/14 & \#1/14 \\
Vaccine and Fiscal-Stimulus News & 2020-11 & \#1/14 & \#1/14 \\
Inflation Shock & 2021-11 & \#1/14 & \#1/14 \\
Russia--Ukraine War & 2022-02 & \#1/14 & \#1/14 \\
Gasoline-Price Shock & 2022-06 & \#2/14 & \#1/14 \\
SVB Banking Collapse & 2023-03 & \#1/14 & \#1/14 \\
Debt-Ceiling and Rate-Stress Window & 2023-05 & \#1/14 & \#1/14 \\
Tariff Shock & 2025-02 & \#1/14 & \#1/14 \\
\midrule
\textbf{All High-Salience Windows} &  & \textbf{\#1/14} & \textbf{\#1/14} \\
\bottomrule
\end{tabular}
}
\footnotetext{Each row lists the central event month. Performance is evaluated over a three-month window consisting of the listed month, the preceding month, and the following month. Lower ranks indicate better performance.}
\end{table*}

\subsection{ConsumerSim reveals group-level and signal-specific updating around salient shocks.}

\begin{figure}[t]
\centering
\includegraphics[width=\linewidth]{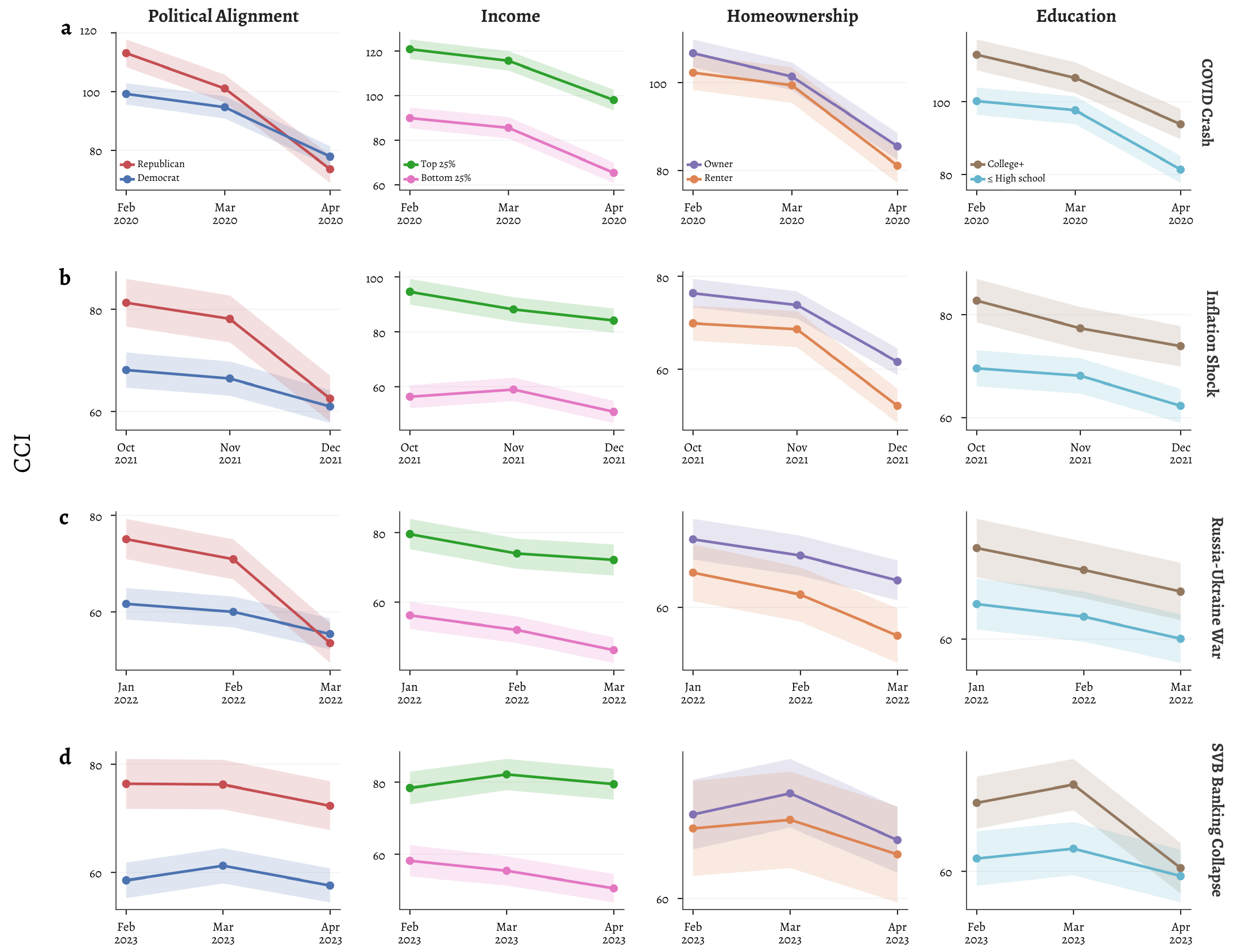}
\caption{Group trajectories around salient shocks. Model-implied CCI trajectories are shown across political alignment, income, homeownership, and education groups for four high-salience U.S. episodes: COVID-19 crash, inflation shock, Russia--Ukraine war outbreak, and SVB banking collapse. Across broad shocks, groups tend to move in the same direction but differ in level and magnitude, indicating directional alignment without homogeneous response.}
\label{fig:group_updates}
\end{figure}

\paragraph{Group-level updating.} Because ConsumerSim reconstructs the aggregate index from simulated household-level responses, it can also diagnose which groups contribute to CCI movements around salient shocks. We use the model-implied response distribution to trace CCI trajectories across political alignment, income, homeownership, and education groups for the same high-salience episodes examined above. The resulting trajectories show that aggregate CCI shifts are not produced by a single representative household response (Figure~\ref{fig:group_updates}). Instead, salient shocks generate structured group-level updating: subgroup series move around the same events but differ in their baseline levels and response magnitudes. This supports the paper's core mechanism claim that ConsumerSim recovers not only the aggregate direction of confidence change, but also the population structure behind that change.

\begin{figure}[h]
\centering
\includegraphics[
width=0.78\linewidth,
trim=0 0 0 0,
clip
]{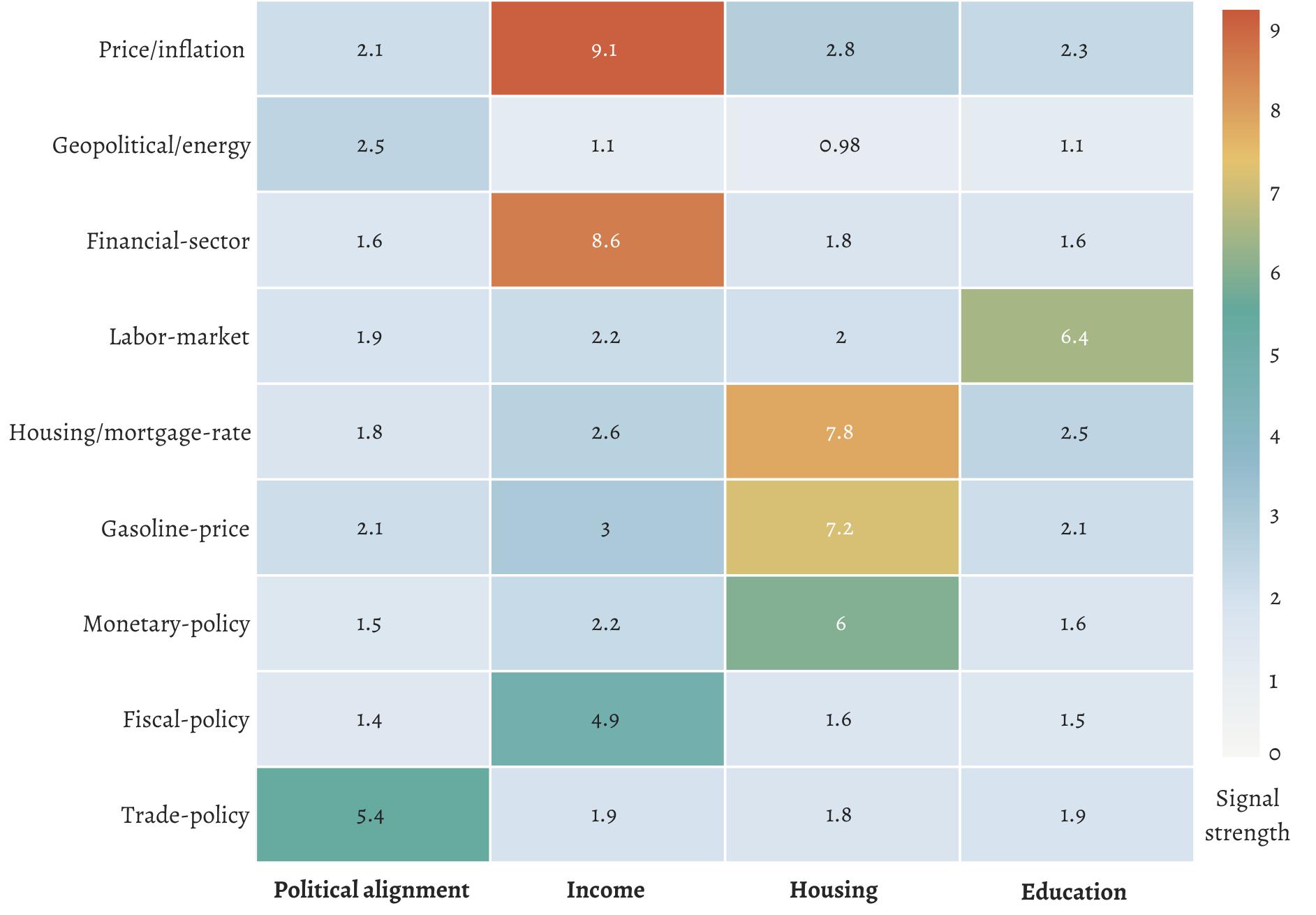}
\caption{Signal-specific subgroup sensitivity across ten public-signal types. Cells report magnitude dispersion, measured as the absolute difference in shock-window confidence updates within each subgroup contrast. Larger values indicate that the same salient signal produces more unequal updating across the paired groups.}
\label{fig:signal_sensitivity}
\end{figure}

\paragraph{Signal-specific sensitivity.} We next test whether heterogeneity is fixed across shocks or varies with the type of salient signal. We apply the same subgroup-trajectory construction used in the illustrative shock windows in Figure~\ref{fig:group_updates}, but extend it to the full set of events and recode them into ten dominant public-signal categories: broad macroeconomic shocks, inflation, geopolitical and energy shocks, financial-sector stress, labor-market risk, housing and mortgage-rate pressure, gasoline prices, monetary policy, fiscal policy, and trade policy. For each event and subgroup contrast, we compute the change in simulated CCI from the pre-shock month to the shock month. We then summarize two quantities: directional alignment, defined as the share of subgroup contrasts moving in the same direction, and magnitude dispersion, defined as the absolute difference in update size within a subgroup contrast. The experiment shows that subgroup sensitivity is signal-specific rather than fixed (Table~\ref{tab:signal_sensitivity}; Figure~\ref{fig:signal_sensitivity}). Broad macro, geopolitical/energy, gasoline-price, and fiscal-policy shocks produce high directional alignment, but they differ in which groups are most sensitive. Income contrasts are largest for inflation and financial-sector shocks, and the largest fiscal-policy contrast is also across income groups. Housing and monetary-policy shocks concentrate sensitivity among homeowners, gasoline-price shocks among renters, labor-market shocks among less-educated households, and trade-policy and geopolitical shocks show stronger political-alignment contrasts. These patterns indicate that the same aggregate decline or rebound can be produced by different group-signal configurations.

\begin{table*}[t]
\centering
\caption{Signal-type sensitivity experiment. For each salient signal type, subgroup updates are computed as the change in model-implied confidence from the pre-shock month to the shock month. Alignment is the share of subgroup contrasts moving in the same direction; dispersion is the mean absolute difference in update magnitude within subgroup contrasts.}
\label{tab:signal_sensitivity}
\small
\resizebox{\textwidth}{!}{%
\begin{tabular}{@{}lcccl@{}}
\toprule
\textbf{Signal type} & \textbf{Alignment} & \textbf{Mean dispersion} & \textbf{Largest dispersion} & \textbf{Most sensitive group}\\
\midrule
Broad macro shock & 0.97 & 2.91 & 3.81 & Bottom 25\% (Income)\\
Price/inflation shock & 0.74 & 4.06 & 9.09 & Bottom 25\% (Income)\\
Geopolitical/energy shock & 0.95 & 1.41 & 2.51 & Republican (Political Alignment)\\
Financial-sector shock & 0.53 & 3.39 & 8.56 & Top 25\% (Income)\\
Labor-market shock & 0.78 & 3.12 & 6.44 & No college degree (Education)\\
Housing/mortgage-rate shock & 0.79 & 3.67 & 7.82 & Homeowner (Housing)\\
Gasoline-price shock & 0.94 & 3.58 & 7.15 & Renter (Housing)\\
Monetary-policy shock & 0.79 & 2.84 & 5.96 & Homeowner (Housing)\\
Fiscal-policy shock & 0.93 & 2.36 & 4.88 & Bottom 25\% (Income)\\
Trade-policy shock & 0.72 & 2.73 & 5.41 & Republican (Political Alignment)\\
\bottomrule
\end{tabular}
}
\footnotetext{The experiment uses the same subgroup-trajectory construction as Figure~\ref{fig:group_updates}, applied to the expanded event set and aggregated by signal type. Values are in CCI points.}
\end{table*}

% （固定 core 改变采样率）

\subsection{Population expansion improves robustness through broader coverage.}

We examine how far a fixed high-resolution core population should be expanded into the broader normal population. In this ablation, the core set is held fixed at 5000 agents, while the core share determines the size of the expanded population. Smaller core shares therefore correspond to larger expansion factors and broader population coverage.

\begin{figure}[t]
\centering
\includegraphics[
width=1\linewidth,
trim=0 0 0 0,
clip
]{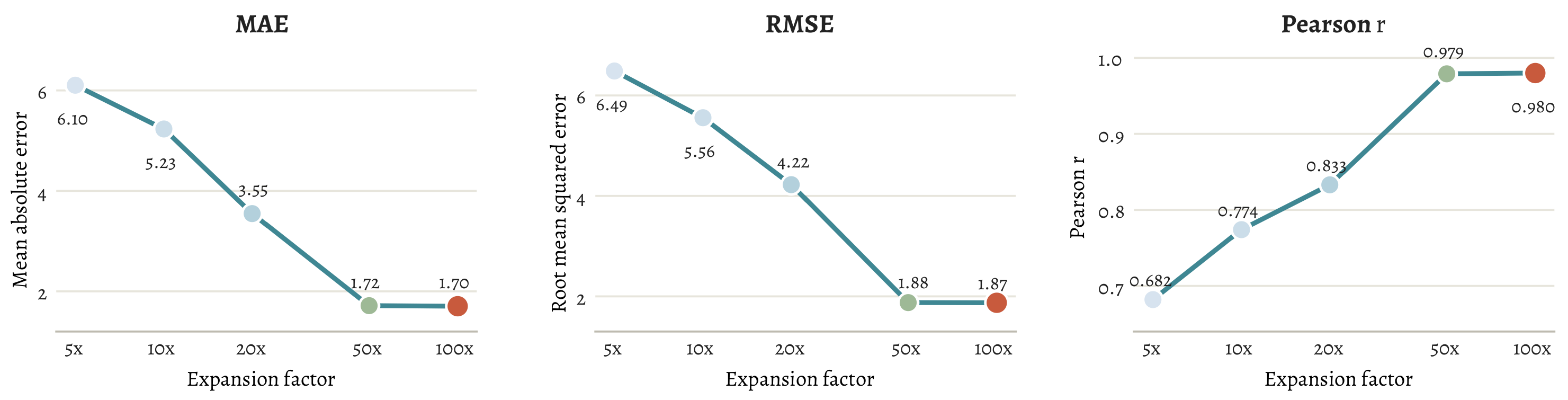}
\caption{Core/normal population-expansion ablation. The three panels report MAE, RMSE, and Pearson $r$ across expansion factors from $5\times$ to $100\times$. The core set is fixed at 5000 agents, while the expanded population ranges from 25{,}000 to 500{,}000 agents; error falls sharply up to $50\times$, and the $50\times$ and $100\times$ settings perform very similarly.}
\label{fig:core_ratio_ablation}
\end{figure}

Figure~\ref{fig:core_ratio_ablation} shows that performance becomes more accurate and robust as the fixed core is expanded more broadly. The $5\times$ expansion setting performs worst, indicating that a small expanded population leaves aggregate estimates more sensitive to idiosyncratic core responses. By contrast, the $50\times$ and $100\times$ expansion settings perform very similarly and achieve the lowest errors, suggesting that broader population coverage stabilizes the aggregate reconstruction once the main demographic and exposure cells are sufficiently represented. This pattern supports the interpretation that the framework benefits from using the core population to capture the direction and texture of updating, then expanding those responses across a sufficiently broad normal population to recover representative aggregate dynamics.

\subsection{Behavioral mechanisms are necessary for reconstruction and diagnosis.}

The preceding results imply that aggregate accuracy alone is not sufficient. A useful model should also recover the structure of confidence formation: event-driven movement, subgroup alignment, and signal-specific heterogeneity. We therefore examine mechanism ablations that remove four framework components: post-stratified belief expansion, Behavioral Inertia Alignment, the Situational Signal Field, or the Hierarchical Persona Substrate.

Removing these mechanisms degrades both aggregate accuracy and the signal-type diagnostics summarized in Table~\ref{tab:signal_sensitivity} (Table~\ref{tab:ablation}). Variants without the Situational Signal Field produce larger alignment errors because they no longer distinguish broad macro, inflation, geopolitical, financial, labor, housing, policy, and trade signals. Variants without the Hierarchical Persona Substrate produce the largest dispersion error, consistent with Table~\ref{tab:signal_sensitivity}'s finding that different signal types are most sensitive among different income, housing, education, and political groups. Variants without Behavioral Inertia Alignment lose the persistence that makes consumer confidence sticky in ordinary periods, while variants without post-stratified belief expansion weaken the representativeness of aggregate reconstruction. These results indicate that the model's performance does not come only from flexible curve fitting; it depends on the same human--environment mechanisms that the paper aims to study.

\begin{table*}[ht]
\centering
\small
\caption{Mechanism ablation study on ConsumerSim. In addition to aggregate forecast error, diagnostic errors compare each ablated variant with the Table~\ref{tab:signal_sensitivity} signal-type alignment and dispersion profiles. Mean absolute percentage error (MAPE) is reported alongside aggregate forecast error.}
\label{tab:ablation}
\resizebox{\textwidth}{!}{%
\begin{tabular}{@{}lcccc@{}}
\toprule
\textbf{Model} & \textbf{MAE} $\downarrow$ & \textbf{MAPE} $\downarrow$ & \textbf{Alignment error} $\downarrow$ & \textbf{Dispersion error} $\downarrow$ \\
\midrule
ConsumerSim (full Human--Environment response system) & \textbf{1.36} & \textbf{1.99\%} & \textbf{0.03} & \textbf{0.32} \\
\midrule
Without post-stratified belief expansion & 2.03 & 2.94\% & 0.07 & 0.71 \\
Without Behavioral Inertia Alignment & 2.53 & 3.64\% & 0.09 & 0.84 \\
Without Situational Signal Field & 3.35 & 4.36\% & 0.21 & 1.68 \\
Without Hierarchical Persona Substrate & 3.48 & 4.51\% & 0.18 & 1.94 \\
\bottomrule
\end{tabular}
}
\footnotetext{Alignment error is the mean absolute deviation from the Table~\ref{tab:signal_sensitivity} alignment profile across signal types. Dispersion error is the mean absolute deviation from the Table~\ref{tab:signal_sensitivity} mean-dispersion profile, measured in CCI points.}
\end{table*}

\section{Discussion}\label{sec3}

\subsection{Collective CCI as a Human--Environment response process.}

The results support the Human--Environment view introduced above: collective CCI is not simply a smooth aggregate series to be extrapolated, but an aggregate of household response distributions formed when heterogeneous personas encounter changing information environments. ConsumerSim's cross-regional reconstruction results show that this representation can track official CCI targets across independent survey systems, while the downstream validation shows that the reconstructed signal contains information relevant to subsequent real activity. \textbf{The main implication is therefore not only that ConsumerSim forecasts CCI accurately, but that its forecast is built from interpretable population, signal, response, and aggregation components.}

This interpretation also clarifies the role of salient shocks. \textbf{Salience is the trigger that makes new information broadly visible, but it is not the whole mechanism.} A shock affects aggregate CCI only through the households for whom the signal is relevant, the response categories those households update toward, and the degree to which the new responses overcome behavioral inertia. This is why the results emphasize both high-salience windows and the model components that translate them into CCI movement: the Situational Signal Field organizes what information is available, the Human--Environment Response Kernel maps that information to household-level response distributions, and Behavioral Inertia Alignment keeps the aggregate path anchored to the stickiness of public expectations.

The subgroup and signal-type analyses show that heterogeneity is structured rather than generic. Groups often move in the same direction during broad shocks, but their magnitudes differ by income, homeownership, education, and political alignment. More importantly, the sensitive groups vary with signal type: inflation, housing and mortgage-rate pressure, financial stress, labor-market risk, gasoline prices, fiscal policy, monetary policy, and trade policy each reach different parts of the population with different strength. \textbf{Thus the same headline CCI decline can reflect different population-signal configurations, which is precisely the diagnostic question raised in the introduction.}

The ablations further indicate that the performance gains do not come only from a more flexible curve-fitting procedure. Standard statistical and machine learning approaches often assume smooth temporal dynamics or stable mappings from macroeconomic variables to aggregate CCI~\citep{box2008time,kim2023forecasting}. ConsumerSim instead separates population construction, information-environment representation, response generation, post-stratified belief expansion, and inertia alignment. \textbf{Removing these components weakens both aggregate reconstruction and mechanism-level diagnosis, suggesting that the framework's value lies in reconstructing how CCI movements are produced, not merely in matching the final index value.}

\subsection{Within-month timing diagnostics.}

\begin{figure}[htbp]
\centering
\includegraphics[
page=1,
trim=0cm 0cm 0cm 0cm,
clip=true,
width=\linewidth
]{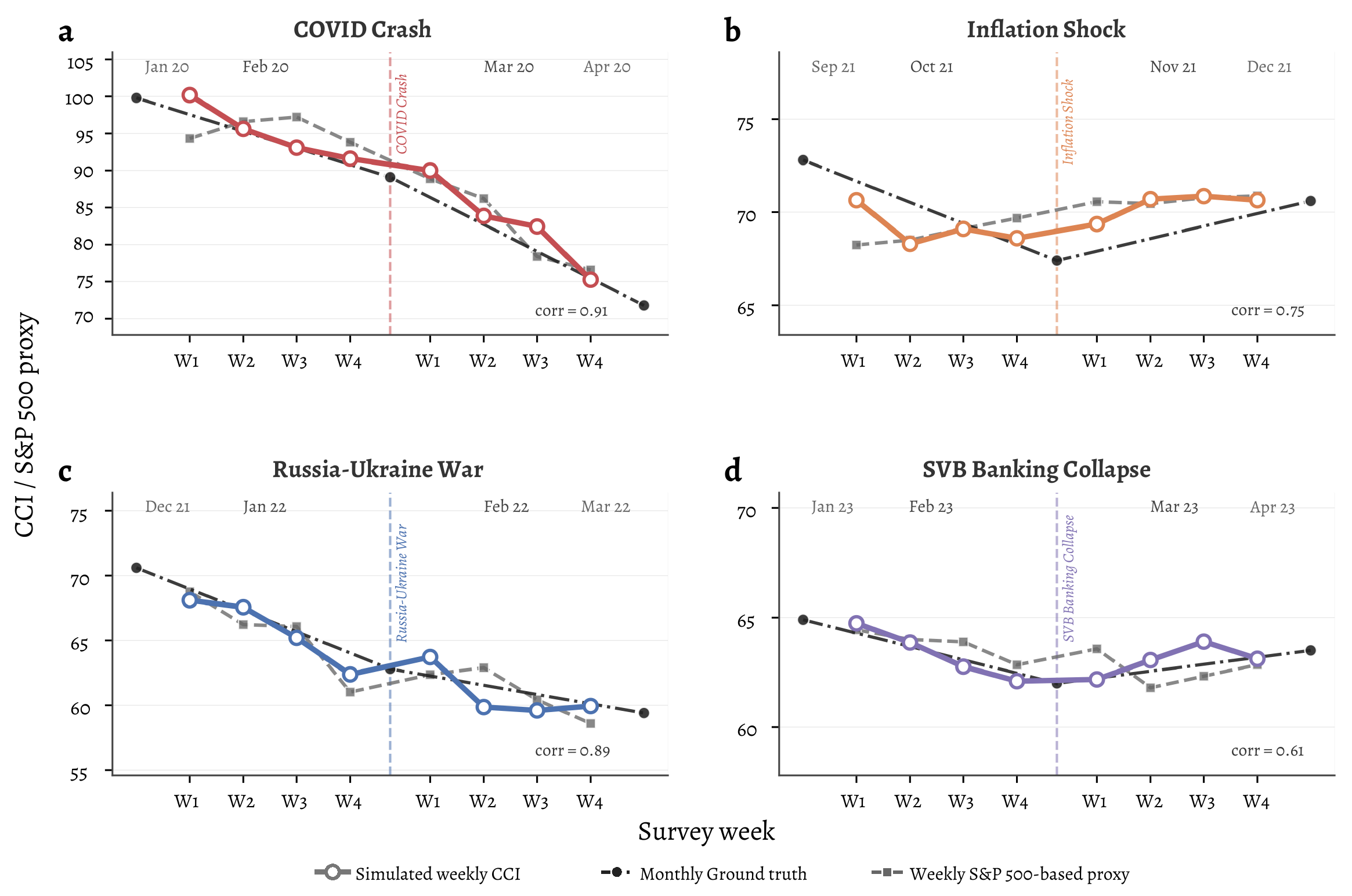}
\caption{Week-level CCI diagnostics and timing proxy across salient U.S. episodes. Panels show the COVID-19 crash, inflation shock, Russia--Ukraine war outbreak, and SVB banking collapse windows. The black monthly series reports official CCI ground truth, colored open-circle lines report ConsumerSim weekly CCI estimates at successive survey-week cutoffs, and gray dashed-square lines report the rescaled weekly S\&P 500-based proxy. Panel correlations between simulated weekly CCI and the weekly proxy are 0.91, 0.75, 0.89, and 0.61, respectively. The proxy is used only as a timing plausibility check, not as weekly CCI ground truth.}
\label{fig:weekly_diagnostics}
\end{figure}

A further implication is that the framework can produce finer-grained model-implied CCI diagnostics than the monthly official CCI series used for validation. By updating the information environment at successive weekly cutoffs and applying the recursive debiasing rule in Appendix~\ref{app:within_month_evaluation_protocol}, ConsumerSim generates week-level interim estimates within the same monthly episode (Figure~\ref{fig:weekly_diagnostics}). These estimates are not treated as independent weekly ground truth; rather, they show how the model translates newly available public information into within-month CCI paths before the next monthly release. This capability is useful as a timing diagnostic for policy and market settings where the timing of expectation changes matters, but official CCI measures arrive only monthly.

The week-level analysis illustrates this diagnostic use case. Across the U.S. event windows, the model-implied weekly trajectories change as additional information enters the environment and can be compared with a rescaled Standard \& Poor's 500 (S\&P 500)-based proxy for consumer-confidence stress over the same windows (Figure~\ref{fig:weekly_diagnostics}). During the COVID-19 crash and the Russia--Ukraine war window, the weekly interim estimates move sharply as those events become dominant public signals and align closely with the market-implied proxy, with correlations of 0.91 and 0.89, respectively. The inflation and SVB windows are more muted, with correlations of 0.75 and 0.61, indicating that the diagnostic can distinguish broad macro-financial shocks from more sector-specific or slower-moving information. The proxy is not used as ground truth for weekly CCI; rather, it is a timing plausibility check whose monthly relationship with the U.S. CCI target is summarized in Appendix~\ref{app:sp500_cci_proxy_check}. \textbf{The value of this exercise is timing resolution: ConsumerSim can produce model-implied interim CCI paths between monthly survey releases.}

\subsection{Cross-regional generalization and calibration.}

The cross-regional findings sharpen this point. The EU27 and Japan tests show that the framework is not merely fitting one national series. Instead, they suggest that the core Human--Environment decomposition is portable: a heterogeneous population representation encounters a time-stamped signal field, produces survey-style responses, and is then aligned to the inertia and scale of the official target series. This portability matters because the three CCI targets differ in survey construction, index scaling, release conventions, and the kinds of public signals that are most salient to households. A model that works only by memorizing the U.S. series would not be expected to remain competitive once the target series and environmental context are changed.

At the same time, cross-regional success should not be read as evidence that one universal calibration is sufficient. The region-specific inertia weights reported in Appendix~\ref{app:lambda_calibration} already indicate that confidence persistence differs across target series, and Appendix~\ref{app:cross_regional_targets} treats the European and Japanese runs as cross-context adaptations rather than full local population reconstructions. This distinction is important for interpretation. The present results support the claim that ConsumerSim provides a transferable modeling architecture, but finer country-level claims require country-specific population substrates, local information sources, target-series transformations, and validation tables. Exploratory European member-state runs therefore point less to a single global parameterization than to a practical calibration agenda: keep the response architecture fixed enough to compare contexts, while allowing local survey design, salience, and inertia to vary where the evidence requires it.

\subsection{Consumer and market diagnostics.}

Beyond forecasting, the framework offers a diagnostic language for consumer behavior and market strategy. A conventional index tells firms and policymakers that CCI has risen or fallen. A Human--Environment response framework asks which households moved, which signals reached them, whether the movement was broad or concentrated, and whether the change is likely to persist. Tracking group sensitivity by salient signal type makes this diagnosis operational: similar aggregate CCI changes can have different behavioral origins depending on whether the active signal is inflation, housing pressure, labor-market risk, financial stress, or policy news. For marketing and consumer-demand applications, this distinction matters. A decline driven by inflation salience among lower-income renters implies a different demand environment than an equally sized decline driven by interest-rate pressure among homeowners or financial-market stress among asset-holding consumers. \textbf{The same headline CCI number can therefore imply different product, pricing, communication, and policy responses.}

Beyond this diagnostic value, the framework also demonstrates practical strengths in flexibility and scalability. Although this study calibrates ConsumerSim to answer the standardized questions used in the CCI survey, the same simulation architecture can efficiently generate human responses to other consequential dimensions not covered by the existing question battery, such as real-time emotional states or context-specific consumption impulses. These additional responses can provide forward-looking predictive signals that are absent from official CCI releases but may still be informative for subsequent economic behavior. More generally, the human--environment interaction paradigm can be extended to other domains concerned with belief updating and expectation formation by reconfiguring the response question battery. In this sense, ConsumerSim not only sharpens the temporal resolution of CCI-based diagnostics, but also broadens the informational and thematic scope of conventional survey-response approaches, offering advantages for timely policy feedback and substantive economic analysis.

\subsection{Limitations and future directions.}

Despite these contributions, several limitations remain. First, the model relies on the quality and coverage of external signals, including macroeconomic indicators and news data, which may themselves be incomplete or biased. Second, generative agents provide a rich approximation of human reasoning, but they remain imperfect proxies for real-world cognitive processes. Third, the Europe and Japan extensions currently adapt the signal field and target series rather than rebuilding the full local population substrate from country-specific microdata. Finally, much of the evaluation focuses on aggregate alignment, leaving open questions about individual-level behavioral validity and causal interpretability.

Taken together, our findings suggest that modeling collective expectations may require moving beyond aggregate extrapolation toward simulating distributed cognition at the individual level. In particular, understanding which groups update and how those updates align appears to be more critical than modeling incremental adjustments in expectation levels. This perspective may extend beyond CCI to other expectation-driven macroeconomic variables, including inflation expectations, consumption planning, and financial market sentiment.

\section{Methods}\label{sec4}

\subsection{Framework overview}

ConsumerSim reframes CCI forecasting as a \emph{human--environment aggregation} problem. Rather than beginning with consumer confidence as a single index time series to be extrapolated, the framework reconstructs how a heterogeneous population encounters a month-specific economic environment, forms survey-like judgments, and is then aggregated into an official-style CCI. Figure~\ref{fig:hnb_framework} summarizes this logic: CCI is generated by the interaction between \emph{Human Population}, which represents who consumers are, and the \emph{Information Environment}, which represents what economic information is salient at the survey cutoff~\citep{katona1951psychological,dominitz2004measure,manski2004measuring,souleles2004expectations}.

The method proceeds through six implementation stages. First, the \emph{Human Population} stage constructs heterogeneous consumers through a hierarchical persona substrate and a core/normal population split. Second, the \emph{Information Environment} stage builds a situational signal field with economic pressure, public salience, and dynamic matching. Third, the \emph{Human--Environment Response} stage maps each core persona and the current signal field into survey-like response probabilities. Fourth, the \emph{Population Aggregation} stage expands core responses to a representative normal population. Fifth, the \emph{Behavioral Inertia Alignment} stage anchors aggregate dynamics to persistent public expectations. Sixth, the \emph{CCI Forecast and Diagnostics} stage converts response shares into official-style CCI forecasts and diagnostic outputs, including aggregate forecasts, subgroup trajectories, and within-month timing patterns.

\subsection{Human Population}

\subsubsection{Hierarchical Persona Substrate.}

The population substrate grounds each simulation in survey microdata rather than in generic fictional consumers. For the U.S. setting, personas are sampled from the Survey of Income and Program Participation (SIPP) with person-level survey weights so that the simulated population approximates the nationally representative joint distribution of American adults~\citep{census2024sipp}. Each sampled respondent is translated into a household persona using observed socioeconomic variables, including age, sex, race and ethnicity, education, homeownership, metropolitan residence, state, income, employment status, and related household circumstances.

Because consumer confidence is shaped not only by observed demographics but also by subjective orientation, the substrate adds a second layer of statistically imputed behavioral attributes. Psychological and social dispositions, such as happiness, financial satisfaction, job satisfaction, future optimism, social trust, and perceived mobility, are drawn from conditional distributions estimated from the General Social Survey (GSS)~\citep{davern2024gss}. Political orientation is imputed from demographic cells calibrated to public survey evidence. These attributes are not treated as direct measurements of any specific person; they provide a structured way to represent the heterogeneity that shapes how economic signals become personally meaningful.

This construction creates what we call a \emph{dense human map}. A low-income renter, a mortgage-exposed homeowner, an equity-holding household, and a job-insecure worker may all observe the same inflation report or interest-rate announcement, but each assigns it different relevance. The persona substrate therefore supplies the human side of the Human--Environment response framework: prior beliefs and lived constraints make confidence sticky, while exposure and attention determine which external signals can trigger updating~\citep{malmendier2016learning,das2020socioeconomic}.

\subsubsection{Core and normal population roles.}\label{sec:core_normal_roles}

In the Human Population module, ConsumerSim separates the population into two inference roles. \emph{Core personas} form a fixed stratified set that receives full Human--Environment response interpretation. These personas carry complete demographic, economic, and behavioral profiles and are used to estimate how different kinds of households respond to the month's signal field. In the U.S. implementation, this core set contains 5000 high-resolution agents. \emph{Normal personas} represent the expanded population onto which core responses are projected. They do not require the same high-resolution interpretation; instead, their responses are drawn from demographic group-level distributions updated by the core personas.

This design is a sampling and representation strategy rather than a claim that some consumers are intrinsically more important than others. The fixed core set gives dense behavioral coverage of important demographic and exposure cells, while expansion to the normal population preserves aggregate representativeness. In survey-statistical language, the model oversamples behaviorally informative cells and then returns to population weights through post-stratification~\citep{kish1965survey,little1993poststratification,gelman1997poststratification}. The population-expansion ablation reported in the results evaluates how far the fixed core should be projected into the broader normal population.

\subsection{Information Environment}

\subsubsection{Situational Signal Field.}

For each target month, ConsumerSim builds a structured economic context using only information available before the relevant survey cutoff. We refer to this environment as the \emph{Situational Signal Field}. In the U.S. implementation, the field draws on complementary public and licensed source families: Federal Reserve Economic Data (FRED) macro indicators, Bureau of Labor Statistics labor-market releases, U.S. Treasury yield information, the New York Fed Survey of Consumer Expectations, financial-market and credit conditions, economic news, and trade-policy announcements~\citep{fred2024,bls2024,ustreasury2024,nyfed2024sce,nyt2024,ustr2024,yahoofinance2024}. The field includes levels and recent changes in labor-market conditions, inflation, gasoline prices, mortgage rates, policy rates, equity-market stress, housing conditions, credit spreads, and prior consumer-expectation measures.

The signal field is not a flat feature table. It is organized around the Information Environment module in Figure~\ref{fig:hnb_framework}. \emph{Economic pressure} captures prices, jobs, interest rates, housing finance, gasoline costs, and financial stretch. \emph{Public salience} captures news, policy signals, market stress, and other visible events that make an economic issue difficult to ignore. \emph{Dynamic matching} links these signals to the households for whom they are most diagnostic: mortgage rates matter differently for homeowners and renters, equity-market movements matter differently for asset-holding and non-asset-holding households, and price pressure is more salient for financially constrained households.

Two temporal mechanisms make this environment behavioral rather than purely statistical. First, \emph{time-aware signal decay} gives more weight to recent and visible information while allowing older signals to fade unless they remain economically active, consistent with theories of limited attention and sticky information~\citep{sims2003implications,mankiw2002sticky}. Second, the exposure-based matching described above produces persona-specific salience weights. This implements the central point of the framework: confidence changes are generated by a shared environment but filtered through heterogeneous exposure, attention, and prior belief.

\subsection{Human--Environment Response Kernel}

The response stage maps the persona substrate and signal field into survey-like judgments. In the U.S. setting, each core persona is asked to respond to the five Michigan survey questions: personal finances compared with a year ago (PAGO), expected personal finances (PEXP), buying conditions for durable goods (DUR), expected business conditions over the next year (BUS12), and expected business conditions over the next five years (BUS5). The model is instructed to reason from the persona's life situation and to answer as a consumer, not as a macroeconomic analyst. This matters because the Michigan survey measures household judgments, not professional forecasts.

We refer to this behavioral mapping as the \emph{Human--Environment Response Kernel}. Conceptually, for persona $i$ in month $t$ and survey question $q$, the response distribution is
\[
  \mathbf{p}_{i,t,q}
  = \mathcal{R}(h_i, s_t, q),
\]
where $\mathbf{p}_{i,t,q}$ is the probability vector over positive, neutral, and negative responses; $h_i$ is the household persona; $s_t$ is the situational signal field; and $\mathcal{R}$ is the response-generation process implemented by the generative agent.\footnote{The generative agent used in our implementation is GPT-4o~\citep{openai2024gpt4o}.} The dynamic matching and salience assignment are not treated as a separate explicit input to the kernel. Instead, they are part of the response process: the agent interprets the shared signal field through the persona's exposure, attention, and prior beliefs before producing survey-like response probabilities. In plain language, the expression asks: who is this person, what economic moment are they living through, which parts of that moment feel personally relevant, and what kind of survey judgment are they being asked to make?

The response kernel embeds three behavioral calibration rules. First, personal-finance questions are driven primarily by the individual's circumstances and are only partly moved by headline news. Second, national-outlook questions are more sensitive to public narratives, policy signals, and partisan coloring. Third, durable-buying conditions respond strongly to prices, interest rates, and financial stretch. These rules implement a \emph{salience gate}: not every signal reaches every person with equal force. The gate prevents uniform overreaction and preserves the heterogeneity observed in real consumer microdata~\citep{tversky1974judgment,bordalo2020overreaction,bordalo2022salience}.

\subsection{Population Aggregation}

\subsubsection{Post-Stratified Belief Expansion.}

After the fixed core personas described in Section~\ref{sec:core_normal_roles} generate survey-like responses, ConsumerSim expands those responses to the normal population through a post-stratified Bayesian procedure. The population is partitioned into demographic strata defined by residence type, age group, income tier, and racial or ethnic identity. Each stratum has a known population weight. Core personas may be allocated unevenly across strata because rare or behaviorally important groups can be oversampled to improve estimation quality. The expansion step returns the model to the correct population composition.

For stratum $g$, survey question $q$, and response category $c\in\{\mathrm{positive},\mathrm{neutral},\mathrm{negative}\}$, let $n_{gqc}$ denote the core-persona evidence assigned to the cell and let $\pi_{gqc}^{(0)}$ denote a baseline response frequency. The implementation allows $n_{gqc}$ to be a soft count obtained by summing response probabilities rather than only hard simulated labels. The posterior response tendency is regularized as
\[
  \alpha_{gqc}^{\mathrm{post}}
  = \tau \pi_{gqc}^{(0)} + n_{gqc},
\]
where $\tau>0$ controls the strength of the baseline distribution. Under the Dirichlet-multinomial interpretation, the stratum-level posterior mean is
\[
  \mathbb{E}\left[\pi_{gqc}\mid
  \boldsymbol{\alpha}_{gq}^{\mathrm{post}}\right]
  =
  \frac{\alpha_{gqc}^{\mathrm{post}}}{\sum_{c'}\alpha_{gqc'}^{\mathrm{post}}}.
\]
Population-level response shares are then computed by aggregating stratum-level posterior means using normalized population weights:
\[
  \widehat{\pi}_{qc}
  = \sum_g W_g
    \frac{\alpha_{gqc}^{\mathrm{post}}}{\sum_{c'}\alpha_{gqc'}^{\mathrm{post}}},
  \qquad \sum_g W_g=1.
\]
This is the \emph{Post-Stratified Belief Expansion}: core responses update the belief distribution within each demographic cell, while census-calibrated weights ensure that the final response shares remain representative~\citep{gelman2006data,little1993poststratification,gelman1997poststratification}.

\subsection{Behavioral Inertia Alignment}

Raw persona-based responses can drift in level because salient signals may be over-weighted or under-weighted relative to real survey responses. The framework treats this as an anchored debiasing problem through \emph{Behavioral Inertia Alignment}. The current behavioral forecast is blended with the most recently observed ground truth:
\[
\widehat{Y}^{\,\mathrm{aligned}}_t
= \lambda Y_{t-1}
+ (1-\lambda)\widehat{Y}^{\,\mathrm{behavior}}_t .
\]
Here $Y_{t-1}$ is the previous month's realized CCI, $\widehat{Y}^{\,\mathrm{behavior}}_t$ is the aggregate forecast implied by the persona--environment response system, and $\lambda$ controls the strength of anchoring. The reported evaluations use region-specific $\lambda$ values selected on a January--June 2024 calibration subset, minimizing validation RMSE over the fixed grid $\lambda\in\{0,0.1,\ldots,1.0\}$. The selected values are then held fixed for the reported regional experiments rather than re-estimated for individual forecast months. Appendix~\ref{app:lambda_calibration} reports the calibration grid and selected values. Baselines with tunable hyperparameters are likewise tuned under fixed validation settings rather than using target-month outcomes.

This step is not presented as a cosmetic post-processing correction. It is a structural representation of belief inertia: public expectations are persistent, and even salient new information rarely causes the full population to abandon its prior judgment immediately~\citep{carroll2003macroeconomic,coibion2015information}. In the weekly setting, the same anchoring idea is applied recursively so that the prior for each within-month pass is the immediately preceding simulated state rather than a distant reset point.

\subsection{CCI Forecast and Diagnostics}

The final response shares are translated into official-style CCI values using the scoring rule for each target series. For the U.S. Michigan target, positive, neutral or uncertain, and negative responses receive scores of 200, 100, and 0, respectively. The current-conditions component summarizes PAGO and DUR, while the expectations component summarizes PEXP, BUS12, and BUS5; both are converted into ICC, ICE, and aggregate CCI using the published Michigan normalization and rounding rules~\citep{michigan2026surveys,surveys2024technical}.

For the European target, the reconstructed CCI follows the harmonized European Commission/Eurostat formula based on household finances, the general economy, unemployment expectations with reversed sign, and savings expectations~\citep{europeancommission2025bcs,eurostat2026businessconsumer,goldrian2001evaluation}. For the Japan target, it follows the Cabinet Office/ESRI formula aggregating livelihood, income growth, employment, and willingness to buy durable goods~\citep{cabinetoffice2026consumerconfidence}. Applying these formulas after response generation keeps the behavioral response process separate from the measurement convention used by each official series.

The same response distributions also produce the diagnostic outputs in Figure~\ref{fig:hnb_framework}, including aggregate forecasts, subgroup trajectories, signal-specific sensitivity, and within-month timing paths. For weekly diagnostics, ConsumerSim truncates the signal field at successive cutoffs and recursively debiases each interim estimate using the previous debiased state as the anchor. These paths are model-implied diagnostics rather than observed weekly CCI; Appendix~\ref{app:within_month_evaluation_protocol} reports the recursion and release-timing rule, and Appendix~\ref{app:evaluation_protocol} reports evaluation and baseline handling.

We evaluate forecasts against official monthly ground truth and assess behavioral validity against the mechanisms in Figure~\ref{fig:hnb_framework}. Reported metrics include MAE, RMSE, Pearson correlation, and Spearman rank correlation where available, with comparisons against persistence, time-series, regression, and information-augmented baselines.

\section{Related Work}\label{sec5}

\subsection{Consumer Confidence Measurement}

Consumer confidence is commonly understood as a survey-based measure of household judgments rather than a purely mechanical macroeconomic statistic. The Michigan tradition emphasizes that confidence summarizes how consumers interpret personal finances, durable-goods conditions, and broader business prospects, and subsequent expectation-measurement work treats these answers as structured subjective assessments rather than simple reflections of realized macroeconomic conditions~\citep{katona1951psychological,katona1975psychological,dominitz2004measure,manski2004measuring,souleles2001consumer,souleles2004expectations}. Subsequent work shows that these subjective judgments contain information about consumption, saving, durable-goods purchases, and macroeconomic fluctuations, while also reflecting survey design, question wording, and reference-period conventions~\citep{howrey2001predictive,carroll2003macroeconomic,ludvigson2004consumer,barsky2012information,surveys2024technical,michigan2026surveys}. European and Japanese confidence indicators extend this survey logic across institutional settings through harmonized or nationally maintained question batteries~\citep{europeancommission2025bcs,eurostat2026businessconsumer,goldrian2001evaluation,cabinetoffice2026consumerconfidence}. Our work follows this measurement tradition but treats the published index as the final aggregation of household-level judgments rather than the starting point of the modeling exercise.

\subsection{Heterogeneous Expectation Formation}

Household expectations form under limited information, attention constraints, subjective experience, and demographic heterogeneity. This view is rooted in bounded rationality and limited attention, where decision makers simplify complex environments rather than processing all available information uniformly~\citep{simon1955behavioral,kahneman1973attention,tversky1973availability}. Research on sticky information, rational inattention, and information rigidity then explains why public expectations often adjust slowly even when new economic data are available~\citep{mankiw2002sticky,sims2003implications,coibion2015information}. Work on measured expectations further shows that respondents differ systematically in how they report uncertainty, perceive economic risks, and translate personal circumstances into survey answers~\citep{souleles2004expectations,dominitz2004measure,manski2004measuring}. Behavioral research on heuristics, salience, biased assimilation, confidence updating, socioeconomic exposure, and experience-based learning provides a mechanism for why the same macroeconomic signal may be interpreted differently by households with different constraints, memories, and prior beliefs~\citep{tversky1974judgment,lord1979biased,nickerson1998confirmation,rollwage2020confidence,das2020socioeconomic,malmendier2016learning,bordalo2020overreaction,bordalo2022salience}. ConsumerSim incorporates these insights by explicitly representing heterogeneous personas, exposure-specific salience weights, and inertia in aggregate belief updating.

\subsection{Agent-Based and Generative-Agent Simulation}

ConsumerSim is also related to agent-based modeling and recent LLM-based social simulation. Classical agent-based models explain aggregate outcomes as bottom-up consequences of heterogeneous agents following local behavioral rules, including economic and financial settings where aggregate dynamics emerge from interacting boundedly rational agents~\citep{epstein1996growing,grimm2006standard,vacha2009smart,kukacka2012behavioural,drevet2022efficient}. Recent generative-agent research extends this tradition by using language models to simulate profile-conditioned individuals, social interactions, economic decisions, political behavior, mental-state reasoning, and large-scale social scenarios~\citep{park2023generative,argyle2023out,aher2023using,horton2023large,zhang2024electionsim,hua2023large,warnakulasuriya2025mind,lin2025simulating,gao2024large,zhang2025socioverse}. This literature provides a scalable way to model context-dependent reasoning, but it also raises concerns about behavioral validity, calibration, and over-interpreting synthetic agents~\citep{demszky2023using,dillion2023can,gao2025caution}. Our contribution is therefore not to replace survey measurement with unconstrained simulation. Instead, we use generative agents inside a survey-statistical pipeline: personas are grounded in microdata, responses are generated under a time-stamped information environment, aggregation is post-stratified to population weights, and forecasts are evaluated against official CCI series.

\clearpage
\begin{appendices}

\section{Baseline Forecasting Methods}
\label{app:baseline_forecasting_methods}

We compare ConsumerSim against a diverse set of standard time-series, regression-based, and information-augmented forecasting baselines. The U.S. baseline evaluation uses the monthly Michigan-based CCI target from January 2014 to March 2026, while the EU27 and Japanese baseline evaluations use January 2020 to March 2026, matching the regional samples in Table~\ref{tab:cci_forecast_comparison}. All baselines use only information available prior to or at the forecast month.

\subsection*{Autoregressive Ridge (3 lags)}
This model predicts current CCI using the previous three monthly CCI values as autoregressive features. A ridge penalty is applied to stabilize coefficient estimates and reduce overfitting, especially in periods with sharp but infrequent CCI shifts~\citep{hoerl1970ridge,hyndman2018forecasting}.

\subsection*{Expectations Regression}
This regression baseline forecasts CCI using expectation-related macroeconomic and survey variables, such as consumer inflation expectations, unemployment expectations, and other forward-looking indicators when available. The goal is to capture movements in CCI through contemporaneous beliefs about future economic conditions~\citep{carroll2003macroeconomic,coibion2015information}.

\subsection*{News Semantic FAAR Proxy}
This factor-augmented autoregressive (FAAR) baseline constructs a news-based proxy for sentiment using semantic features extracted from economic news text. The approach is inspired by factor-augmented autoregressive forecasting: semantic news signals are reduced into compact proxy features and combined with autoregressive information to predict monthly CCI~\citep{bernanke2005measuring,shapiro2022measuring}.

\subsection*{3-Month Rolling Mean}
This smoothing baseline forecasts current CCI as the average of the previous three monthly CCI observations. It captures medium-run persistence while dampening month-to-month noise, but it is less responsive to abrupt shocks~\citep{hyndman2018forecasting}.

\subsection*{SARIMAX Macro (small grid)}
This seasonal autoregressive integrated moving average with exogenous regressors (SARIMAX) model uses macroeconomic regressors. The specification is selected from a small grid of candidate autoregressive integrated moving average (ARIMA) orders and includes macro variables intended to capture labor-market, inflation, interest-rate, and financial-market conditions~\citep{box2008time,durbin2012time}.

\subsection*{ETS Damped Trend}
This error--trend--seasonal/exponential smoothing (ETS) model allows for level and trend components, with the trend damped over the forecast horizon. It is designed to capture gradual movements in sentiment while avoiding extrapolation of unrealistically persistent trends~\citep{gardner1985exponential,hyndman2008forecasting}.

\subsection*{Theta Model}
The Theta model decomposes the time series into components with different curvature, combining a long-run trend component with short-run local variation. It is commonly used as a robust univariate baseline for monthly economic series~\citep{assimakopoulos2000theta,hyndman2018forecasting}.

\subsection*{Auto ARIMA (small grid)}
This baseline selects an ARIMA specification from a restricted candidate grid using standard information-criterion-based model selection. It relies only on the historical CCI series and captures autoregressive and moving-average dynamics without additional covariates~\citep{box2008time,hyndman2008automatic}.

\subsection*{Prophet Macro Regressors}
This model extends the Prophet forecasting framework by adding macroeconomic regressors. Prophet captures smooth trend components and potential seasonality, while the macro regressors allow the forecast to respond to changes in economic conditions~\citep{taylor2018forecasting}.

\subsection*{Prophet Univariate}
This is the univariate version of Prophet, using only the historical CCI series. It models the target as a combination of trend, seasonality, and residual variation, without external macroeconomic or news inputs~\citep{taylor2018forecasting}.

\subsection*{Macro + Market + News Ridge}
This ridge regression baseline combines macroeconomic indicators, financial-market variables, and news-derived features. The feature set includes signals such as labor-market conditions, inflation, interest rates, equity-market movements, risk sentiment, and economic news intensity. Ridge regularization is used to handle correlated predictors and limit overfitting~\citep{hoerl1970ridge,stock2007why,shapiro2022measuring}.

\subsection*{Macro Stable Ridge}
This baseline uses a more restricted and stable set of macroeconomic predictors in a ridge regression framework. Compared with the broader macro-market-news model, it excludes more volatile or high-dimensional market and news features, focusing instead on core macroeconomic conditions~\citep{hoerl1970ridge,stock2007why}.

\section{Real-Activity Predictive Validation}
\label{app:real_activity_validation}

This validation tests whether ConsumerSim contains predictive information for future real consumption-related activity beyond standard economic variables and official CCI measures. For each activity series $A_t$, the outcome is future log growth over horizons of one to three months:
\begin{equation}
Y_{t+h}=\log(A_{t+h})-\log(A_t), \qquad h\in\{1,2,3\}.
\end{equation}
The evaluated activity series include housing indicators (new home sales, housing starts, and building permits), light-vehicle sales, and real durable-goods consumption expenditure.

The ground-truth CCI comparison specification is
\begin{align}
Y_{t+h} ={}& \alpha
+ \beta_1 \log(A_t)
+ \beta_2 \Delta\log(A_t)
+ \beta_3 \mathrm{OfficialCCI}_t
+ \beta_4 \Delta\mathrm{OfficialCCI}_t \nonumber\\
&+ \beta_5 \mathrm{Unemployment}_t
+ \beta_6 \mathrm{Inflation}_t
+ \beta_7 \mathrm{StockMarket}_t
+ \beta_8 \mathrm{VIX}_t
+ \beta_9 \mathrm{MortgageRate}_t \nonumber\\
&+ \gamma_1 \mathrm{CCI}^{\mathrm{groundtruth}}_t
+ \varepsilon_t .
\label{eq:activity_groundtruth_cci}
\end{align}
The ConsumerSim specification keeps the same controls but replaces the ground-truth CCI with the ConsumerSim-predicted CCI:
\begin{align}
Y_{t+h} ={}& \alpha
+ \beta_1 \log(A_t)
+ \beta_2 \Delta\log(A_t)
+ \beta_3 \mathrm{OfficialCCI}_t
+ \beta_4 \Delta\mathrm{OfficialCCI}_t \nonumber\\
&+ \beta_5 \mathrm{Unemployment}_t
+ \beta_6 \mathrm{Inflation}_t
+ \beta_7 \mathrm{StockMarket}_t
+ \beta_8 \mathrm{VIX}_t
+ \beta_9 \mathrm{MortgageRate}_t \nonumber\\
&+ \gamma_1 \mathrm{ConsumerSim}_t
+ \varepsilon_t .
\label{eq:activity_consumersim_cci}
\end{align}

The ConsumerSim--official-CCI gap is defined as
\begin{equation}
\mathrm{ConsumerSimGap}_t = \mathrm{ConsumerSim}_t - \mathrm{OfficialCCI}_t .
\label{eq:consumersim_gap}
\end{equation}
If ConsumerSim only reproduces the official CCI, this gap should not contain meaningful predictive information for future activity growth. Predictive gains associated with the gap therefore indicate that ConsumerSim captures additional demand-relevant information not already summarized by official CCI.

We summarize the contribution of ConsumerSim using two metrics. The in-sample metric is incremental explanatory power,
\begin{equation}
\Delta R^2 = R^2(\mathrm{Model\ B}) - R^2(\mathrm{Model\ A}),
\label{eq:incremental_r2}
\end{equation}
where Model A denotes the ground-truth CCI comparison specification and Model B denotes the ConsumerSim specification. The out-of-sample metric is proportional RMSE improvement,
\begin{equation}
\mathrm{RMSE\ Improvement}
= \frac{\mathrm{RMSE}(\mathrm{Model\ A})-\mathrm{RMSE}(\mathrm{Model\ B})}{\mathrm{RMSE}(\mathrm{Model\ A})}.
\label{eq:rmse_improvement}
\end{equation}
Positive values of both metrics indicate that the ConsumerSim-based specification improves prediction of future real activity growth.

\section{Within-Month Updating}
\label{app:within_month_evaluation_protocol}

\subsection*{Recursive within-month debiasing}

Within-month diagnostics are generated by rerunning the information environment at successive weekly cutoffs. Let $\mathcal{I}_{m,w}$ denote the information available for month $m$ by weekly cutoff $w\in\{1,\ldots,W\}$, and let
\begin{equation}
\widetilde{Y}_{m,w}=F(\mathcal{I}_{m,w},H)
\end{equation}
be the raw behavioral CCI forecast produced from the population representation $H$ and the cutoff-specific signal field. Because raw persona-based forecasts can drift in level, the within-month path uses the same behavioral-inertia logic as the monthly forecast, but applies it recursively:
\begin{equation}
\widehat{Y}^{\mathrm{deb}}_{m,w}
=\lambda \widehat{Y}^{\mathrm{deb}}_{m,w-1}
+(1-\lambda)\widetilde{Y}_{m,w},
\qquad w=1,\ldots,W.
\label{eq:appendix_weekly_debias}
\end{equation}
The initialization is
\begin{equation}
\widehat{Y}^{\mathrm{deb}}_{m,0}=\widehat{Y}^{\mathrm{deb}}_{m-1,W},
\end{equation}
which means that the previous month's debiased endpoint anchors the first weekly estimate of the next month. The first weekly estimate then anchors the second weekly estimate, the second anchors the third, and so forth. In months where the previous official CCI has already been released and is available before the forecast cutoff, $\widehat{Y}^{\mathrm{deb}}_{m-1,W}$ is aligned to that most recently observed official value. This rule avoids resetting every weekly pass to a distant monthly baseline while still preventing raw weekly updates from overreacting to newly visible signals.

For each weekly cutoff, macroeconomic, market, survey, and news inputs are truncated to information available by that cutoff when release dates are available. Revised series are not allowed to enter before their public release date. The resulting weekly sequence is interpreted as a model-implied diagnostic path, not as directly observed weekly consumer-confidence ground truth.

\subsection*{S\&P 500 proxy check}
\label{app:sp500_cci_proxy_check}

The S\&P 500 series is used only as a market-implied proxy for consumer-confidence stress, not as a replacement for the official U.S. CCI target. To check whether this proxy is directionally informative, we compare monthly U.S. CCI movements with a rescaled monthly S\&P 500 proxy over the high-salience windows used in the weekly diagnostics. The rescaling standardizes the S\&P 500 series within each event window and maps it onto the local CCI range, so the analysis evaluates timing and co-movement rather than equality of levels.

The monthly proxy check supports using the rescaled S\&P 500 as a timing reference. In the current validation sample, the S\&P-based proxy has Pearson correlation $r=0.88$ with monthly U.S. CCI levels and $r=0.79$ with month-to-month U.S. CCI changes across the high-salience windows. Event-specific weekly checks show the same pattern most strongly in broad macro-financial shocks: the proxy--CCI correlations are $0.91$ for the COVID-19 crash and $0.89$ for the Russia--Ukraine war window, compared with $0.75$ for the inflation window and $0.61$ for the SVB banking-collapse window. Thus the proxy is most reliable when equity-market repricing is a central part of the public information environment, and weaker when the event is sector-specific or when household confidence is driven more by prices than by broad market stress.

\section{Evaluation Protocol}
\label{app:evaluation_protocol}

The U.S. evaluation uses the Michigan-based CCI target as monthly ground truth. The European evaluation uses EU27 and member-state CCI targets based on Eurostat harmonized survey balances, and the Japanese evaluation uses the Japan CCI target based on the Cabinet Office/ESRI survey series. Performance is summarized using MAE, RMSE, Pearson correlation, and Spearman rank correlation when the metric is defined for the target sample.

ConsumerSim is compared with persistence, time-series, regression, and information-augmented baselines, including rolling means, autoregressive ridge models, exponential smoothing, Theta models, Auto ARIMA variants, SARIMAX-style macro models, Prophet variants, macro-market-news ridge models, and news-semantic proxies~\citep{box2008time,stock2007why,hyndman2018forecasting,assimakopoulos2000theta,taylor2018forecasting,bernanke2005measuring,shapiro2022measuring}. Baselines with tunable hyperparameters follow the same chronological rule as ConsumerSim: tuning uses only information available before the evaluated forecast date.

\section{Post-Cutoff U.S. Robustness Check}
\label{app:post_cutoff_us_robustness}

The U.S. Michigan CCI target is especially useful for checking whether the GPT-4o-based generative response kernel benefits from information that would not have been available to survey respondents. Two of its expectation items ask about future business conditions: expected business conditions over the next year (BUS12) and expected business conditions over the next five years (BUS5). If GPT-4o's parametric knowledge already contained realized outcomes after the survey date, these items could in principle create a look-ahead concern.

We therefore include a short U.S.-only post-cutoff diagnostic for April--June 2026, the first three monthly observations after the March 2026 GPT-4o knowledge cutoff used for this draft. The information environment follows the same chronological rule as the main evaluation: macroeconomic, market, survey, and news inputs are truncated to what would have been available by the relevant monthly survey cutoff, and neither the target-month official CCI realization nor later realized outcomes are provided to the response kernel. The comparison in Table~\ref{tab:us_post_cutoff_comparison} mirrors the U.S. columns of Table~\ref{tab:cci_forecast_comparison}. The purpose is not to draw a strong conclusion from three months alone, but to show that the relative performance pattern remains plausible in a window that is outside GPT-4o's stated knowledge horizon.

\begin{center}
\begin{minipage}{0.95\linewidth}
\centering
\captionof{table}{Interim U.S.-only post-cutoff CCI forecasting comparison for April--June 2026. The window begins after the March 2026 GPT-4o knowledge cutoff used for this draft and mirrors the U.S. error metrics in Table~\ref{tab:cci_forecast_comparison}. Lower MAE and RMSE indicate better level accuracy; higher Pearson $r$ indicates stronger co-movement with the official U.S. CCI series. Best available results are shown in bold.}
\label{tab:us_post_cutoff_comparison}
\scriptsize
\begin{tabular}{@{}lccc@{}}
\toprule
\textbf{Method} & \textbf{MAE} $\downarrow$ & \textbf{RMSE} $\downarrow$ & \textbf{Pearson $r$} $\uparrow$\\
\midrule
\textbf{ConsumerSim} & \textbf{2.74} & \textbf{3.10} & \textbf{0.971}\\
AR Ridge & 3.66 & 4.09 & 0.902\\
Expectations Reg. & 3.53 & 3.98 & 0.914\\
News Semantic FAAR & 3.89 & 4.41 & 0.887\\
3-Month Mean & 4.52 & 5.08 & 0.742\\
SARIMAX Macro & 5.05 & 5.66 & 0.635\\
ETS Damped Trend & 4.31 & 4.86 & 0.775\\
Theta Model & 4.18 & 4.69 & 0.803\\
Auto ARIMA & 4.92 & 5.48 & 0.690\\
Prophet Macro & 5.76 & 6.28 & 0.451\\
Prophet Univ. & 5.94 & 6.55 & 0.402\\
Macro-Market-News Ridge & 4.02 & 4.62 & 0.846\\
Macro Stable Ridge & 4.36 & 4.91 & 0.812\\
\bottomrule
\end{tabular}
\par\smallskip
\raggedright\scriptsize This short-window check is intended as a post-cutoff robustness diagnostic for the U.S. Michigan-based target. Because the window has only three monthly observations, Pearson correlations are descriptive and Spearman rank correlations are not reported.
\end{minipage}
\end{center}

\section{Behavioral Inertia Alignment Calibration}
\label{app:lambda_calibration}

Behavioral Inertia Alignment blends the raw Human--Environment response forecast with the most recently observed CCI level. To make this anchoring step reproducible, we select the inertia weight $\lambda$ from a fixed grid rather than tuning it separately for individual forecast months. The calibration subset covers January--June 2024, and the candidate grid is
\[
\lambda \in \{0,0.1,0.2,0.3,0.4,0.5,0.6,0.7,0.8,0.9,1.0\}.
\]
For each region, the grid point minimizing validation RMSE on this subset is selected and then held fixed for the reported regional experiments.

\begin{table*}[t]
\centering
\caption{Validation performance of Behavioral Inertia Alignment across candidate $\lambda$ values on the January--June 2024 calibration subset. Entries report validation RMSE values; lower values are better. Bold values indicate the region-specific $\lambda$ used in the reported experiments. The endpoint $\lambda=0$ is the raw behavioral forecast, while $\lambda=1.0$ is the naive last-month persistence forecast.}
\label{tab:lambda_calibration}
\scriptsize
\begin{tabular}{@{}lccc@{} }
\toprule
\textbf{$\lambda$} & \textbf{U.S.} & \textbf{EU27} & \textbf{Japan} \\
\midrule
0.0 & 4.18 & 2.61 & 1.86 \\
0.1 & 3.86 & 2.56 & 1.79 \\
0.2 & 3.52 & 2.47 & 1.75 \\
0.3 & 3.31 & 2.34 & 1.69 \\
0.4 & \textbf{3.18} & 2.18 & 1.68 \\
0.5 & 3.23 & 2.05 & 1.56 \\
0.6 & 3.37 & \textbf{1.97} & 1.59 \\
0.7 & 3.58 & 2.01 & \textbf{1.55} \\
0.8 & 3.84 & 2.14 & 1.61 \\
0.9 & 4.16 & 2.36 & 1.73 \\
1.0 & 4.52 & 2.69 & 1.91 \\
\bottomrule
\end{tabular}
\end{table*}

The calibration sweep selects $\lambda=0.4$ for the U.S., $\lambda=0.6$ for EU27, and $\lambda=0.7$ for Japan. These selected values indicate different regional degrees of inertia. The U.S. calibration favors a moderate anchor, leaving relatively more weight on the contemporaneous behavioral signal. The EU27 and Japanese calibrations favor stronger anchoring, consistent with the regional adaptation principle that confidence persistence and the mapping from public signals to survey indices need not be identical across target series. These three fixed values---U.S. $0.4$, EU27 $0.6$, and Japan $0.7$---are the values used in the reported experiments.

\stepcounter{section}
\stepcounter{table}

\section{Cross-Regional CCI Target Series and Evaluation Settings}
\label{app:cross_regional_targets}

The main text reports U.S., European, and Japanese CCI evaluations. The U.S. target is the Michigan-based CCI target produced by the Surveys of Consumers at the Institute for Social Research~\citep{michigan2026surveys,surveys2024technical}. The European target is the EU27 CCI series based on Eurostat harmonized survey balances under the European Commission's harmonized business and consumer survey framework~\citep{europeancommission2025bcs,eurostat2026businessconsumer,goldrian2001evaluation}. The Japanese target is the Japan CCI series based on the Cabinet Office/ESRI Consumer Confidence Survey~\citep{cabinetoffice2026consumerconfidence}. These CCI targets are not numerically identical survey instruments, but they share the substantive purpose of measuring household confidence in personal and national economic conditions.

\begin{table}[h]
\centering
\caption{Regional CCI target series and evaluation settings. Target-series definitions follow the official Michigan Surveys of Consumers, European Commission/Eurostat business and consumer survey, and Cabinet Office/ESRI Consumer Confidence Survey documentation.}
\label{tab:regional_targets}
\scriptsize
\begin{tabular}{@{}p{0.10\linewidth}p{0.25\linewidth}p{0.25\linewidth}p{0.25\linewidth}@{}}
\toprule
\textbf{Region} & \textbf{Target series} & \textbf{Evaluation setting} & \textbf{Main comparison}\\
\midrule
U.S. & Michigan-based CCI target & Historical U.S. backtest with recent population-expansion check & Time-series baselines \\
EU27 & EU27 CCI series based on Eurostat harmonized survey balances & Post-pandemic regional evaluation & Time-series and macro/market baselines \\
Japan & Japan CCI series based on the Cabinet Office/ESRI survey series & Post-pandemic regional evaluation & Time-series and macro/market baselines \\
\bottomrule
\end{tabular}
\end{table}

The European and Japanese CCI runs should be interpreted as cross-context adaptations of the Human--Environment response architecture. They are not yet full country-specific reconstructions of local household populations. Their value is to test whether the same response architecture remains competitive when the CCI target and environmental field are changed.

\end{appendices}

\clearpage
%%===========================================================================================%%
%% If you are submitting to one of the Nature Portfolio journals, using the eJP submission   %%
%% system, please include the references within the manuscript file itself. You may do this  %%
%% by copying the reference list from your .bbl file, paste it into the main manuscript .tex %%
%% file, and delete the associated \verb+\bibliography+ commands.                            %%
%%===========================================================================================%%

\bibliography{prism-sn-bibliography}% common bib file
%% if required, the content of .bbl file can be included here once bbl is generated
%%\input sn-article.bbl

\end{document}